\documentclass[12pt]{article}
\usepackage{amsmath}
\usepackage{graphicx}
\usepackage{times}
\usepackage[T1]{fontenc}
\usepackage[utf8]{inputenc}
\usepackage[margin=1cm]{caption}

\usepackage[square,numbers,sort]{natbib}

\author
{Pankhuri Gupta$^{1\dag}$, Kacho Imtiyaz Ali Khan$^{1\dag, \ddag}$, Akash Kumar$^{2,3}$,
\\Rekha Agarwal$^{1}$, Nidhi Kandwal$^{1}$, Ram Singh Yadav$^{1}$,
\\Johan \AA kerman$^{2,3\ast}$ and Pranaba Kishor Muduli$^{1\ast}$
\\
\normalsize{$^{1}$Department of Physics, Indian Institute of Technology Delhi, Hauz Khas,}\\
\normalsize{110016, New Delhi, India}\\
\normalsize{$^{2}$Department of Physics, University of Gothenburg, Fysikgränd 3,}\\
\normalsize{412 96, Gothenburg, Sweden}\\
\normalsize{$^{3}$Research Institute of Electrical Communication, Tohoku University, 2-1-1 Katahira,}\\
\normalsize{Aoba-ku, Sendai 980-8577 Japan}\\
\\
\normalsize{$^\dag$These authors contributed equally to this work.} \\
\normalsize{$^\ddag$Present address: Paul-Drude-Institut für Festkörperelektronik,}\\
\normalsize{Leibniz-Institut im Forschungsverbund Berlin e.V., 10117 Berlin, Germany}\\
\normalsize{$^\ast$To whom correspondence should be addressed; E-mails:} \\
\normalsize{muduli@physics.iitd.ac.in, johan.akerman@physics.gu.se}
}

\date{}

\begin{document}

\parindent 0cm
\parskip 12pt

\title{\LARGE\bfseries{Symmetry Enhanced Unconventional Spin Current Anisotropy in a Collinear Antiferromagnet}} 

\maketitle

\begin{abstract}

Spin-orbit torque (SOT) presents a promising avenue for energy-efficient spintronics devices, surpassing the limitations of spin transfer torque. While extensively studied in heavy metals, SOT in antiferromagnetic quantum materials remains largely unexplored. Here, we investigate SOT in epitaxial FeSn, a collinear antiferromagnet with a kagome lattice. FeSn exhibits intriguing topological quantum features, including two-dimensional flat bands and Dirac-like surface states, making it an ideal platform for investigating emergent SOT properties. Using spin-torque ferromagnetic resonance, we uncover a six-fold symmetric damping-like SOT in epitaxial-FeSn/Py heterostructures, reflecting the six-fold symmetry of the epitaxial [0001]-oriented FeSn films. Additionally, we observe a substantial unconventional field-like torque, originating from spin currents with out-of-plane spin polarization. This torque exhibits a unique angular dependence—a superposition of six-fold crystalline symmetry and uniaxial symmetry associated with the antiferromagnetic spin Hall effect. Notably, the unconventional field-like torque is enhanced when the RF current flows along the Ne\'el vector in FeSn. Our findings reveal an unconventional spin current anisotropy tunable by crystalline and magnetic symmetry, offering a novel approach for controlling SOT in antiferromagnetic spintronics.

\end{abstract}

Spin-orbit torques (SOTs) have emerged as a powerful tool for a wide range of spintronic applications, including ultrafast magnetization switching~\cite{Miron2011Nature,Liu2012,garello2014ultrafast}, spintronic oscillators~\cite{VEDemidov2012,awad2016natphys,kumar2025spinwaves}, and the emerging field of spintronic-based neuromorphic computing~\cite{grollier2020neuromorphic,houshang2020spin, garg2021kuramoto,yadav2023demonstration}. In commonly employed ferromagnet/heavy metal (FM/HM) heterostructures, SOTs are generated either through the spin Hall effect~\cite{Dyakonov1971, Hirsch1999} from the HM or the Rashba-Edelstein effect~\cite{bychkov1984properties,edelstein1990spin} from the interface. 
In both cases, a charge current flowing in the $x-$direction produces a transverse spin current in the $z-$direction with spin polarization along the $y-$direction, a symmetry commonly referred to as ``conventional SOT", with \emph{e.g.} the widely used material Pt and its isotropic SHE the most typical example~\cite{freimuth2010anisotropic, chudnovsky2009intrinsic}. Here, isotropic SHE refers to the fact that the spin Hall conductivity is independent of the relative orientation of the current direction with respect to the crystallographic axis. Conventional SOT in ferromagnetic/heavy-metal heterostructures generate an in-plane anti-damping SOT, which cannot produce deterministic switching of perpendicularly magnetized systems used in high-density magnetic memory.
~\cite{Miron2011Nature, Liu2012,yu2014switching, liu2012current,fukami2016spin}. 

Recent advancements in the field of quantum materials have led to the discovery of novel SOTs in two-dimensional (2D) materials and non-collinear antiferromagnets~\cite{tian2021two, song2021spin}. These materials possess unique symmetry properties that facilitate the generation of unconventional SOT, characterized by spin polarization components along all three spatial axes, irrespective of the current direction. It has been demonstrated that symmetry breaking in 2D quantum materials such as WTe$_2$ ~\cite{macneill2017thickness, Macneil2017NP_WTe2}, and MoTe$_2$~\cite{stiehl2019layer} enables the emergence of such unconventional SOT. Furthermore, non-collinear kagome antiferromagnets like  Mn$_{3}$GaN~\cite{nan2020controlling}, Mn$_{3}$SnN~\cite{you2021cluster}, and Mn$_{3}$Sn~\cite{kondou2021giant},  offer an alternative platform for realizing unconventional SOTs due to their low magnetic symmetry. For all these systems, SOT is intrinsically anisotropic due to its dependence on crystalline and/or magnetic symmetry. 
Recent studies have shown that similar to non-collinear kagome antiferromagnets, collinear kagome antiferromagnetic materials such as Mn$_{2}$Au, CuMnAs, and RuO$_{2}$ have the potential to generate out-of-plane spin polarization due to the broken space-reversal symmetries of the spin sublattice~\cite{chen2021observation, prlRuO2}.

\begin{figure} [t!]
\centering
\includegraphics[width=0.8\columnwidth]{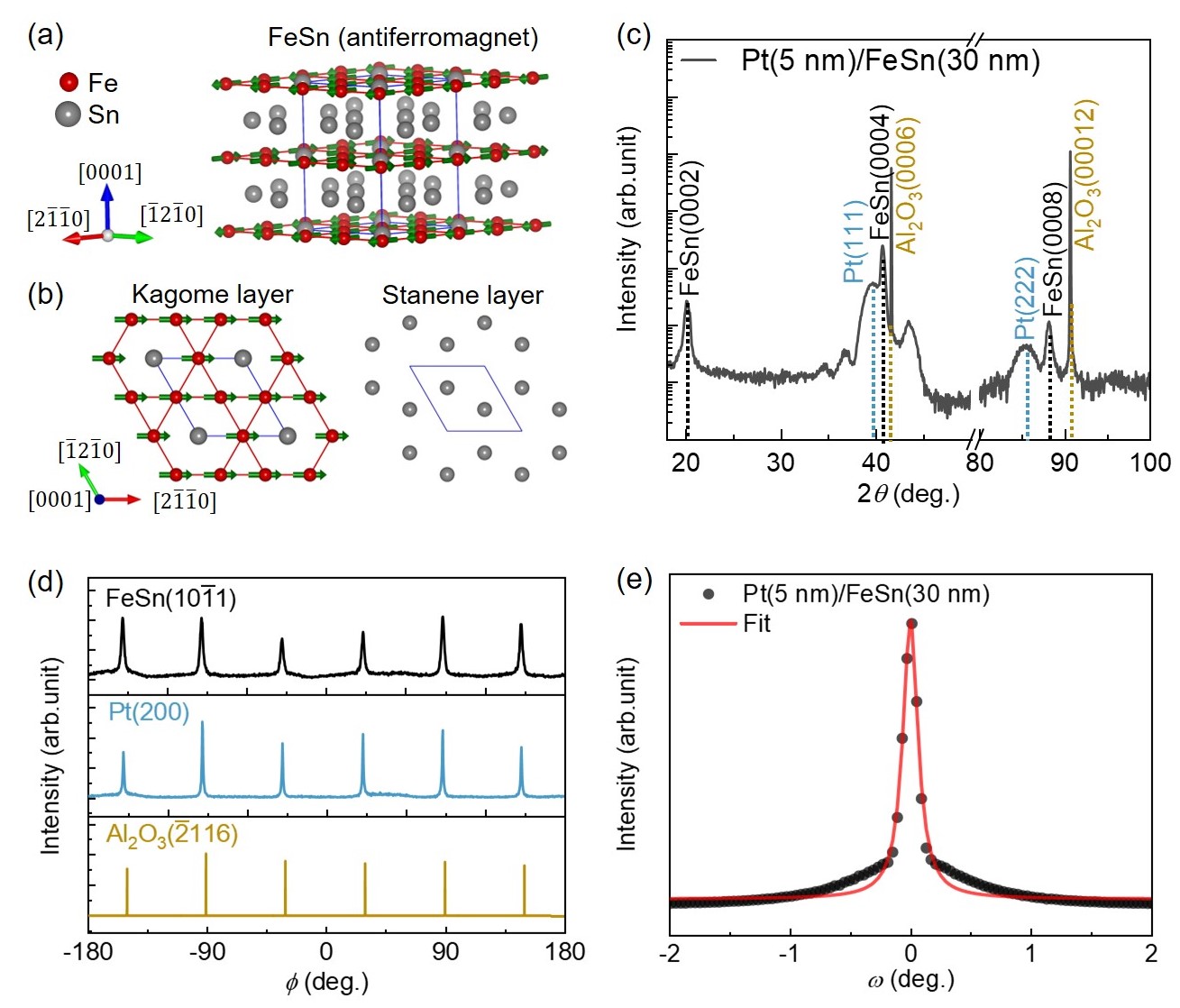}
\caption{\textbf{Schematic of FeSn and  XRD measurements} (a) Schematic of the unit cell (blue) of the kagome antiferromagnet FeSn. (b) Schematic of the kagome layer of Fe$_3$Sn and stanene layer of Sn$_2$. Fe atoms are depicted in red, and Sn atoms are shown in Gray. (c) Structural characterization using $\theta-2\theta$ XRD scan measured for Pt(5 nm)/FeSn(30 nm). (d)~$\phi$-scan measurements for Pt(5 nm)/FeSn(30 nm) film on an Al$_2$O$_3$ substrate. The top, middle, and bottom panels show the $\phi$-scans for FeSn($10\bar{1}1$), Pt(200), and Al$_2$O$_3$($\bar{2}116$) reflections, respectively. (e) The rocking curve ($\omega$-scan) for the Pt(5 nm)/FeSn(30 nm). The black symbol and red line represent the measured data and fitted line to the $\omega$-scan, respectively.}
\label{Fig:1}
\end{figure}

 Kagome antiferromagnetic materials constitute a distinctive class of quantum materials that have garnered significant attention due to their potential to host massless Dirac states~\cite{chowdhury2023kagome}. These topological electronic states arise from the unique kagome lattice structure and the strong spin-orbit coupling~\cite{lin2020dirac, lin2018flatbands, bangar2023large}.
FeSn is a collinear kagome antiferromagnet with a hexagonal structure (space group: P6/mmm) and a Ne\'el temperature of 368~K~\cite{haggstrom1975studies, lin2020dirac, sales2019electronic}. The crystal structure of FeSn [as shown in Fig.~\ref{Fig:1}a$\&$b] consists of alternating kagome layers of Fe$_3$Sn and stanene layers of Sn$_2$, which is slightly different from its kagome ferromagnet Fe$_3$Sn$_2$, where two kagome layers are sandwiched between the stanene layers~\cite{khan2022intrinsic,khan2024magnetodynamic}. However, within each kagome layer of FeSn, Fe atoms align ferromagnetically, while each kagome layer is antiferromagnetically coupled with each other via the stanene Sn$_2$ layer along the \textit{c}-axis. Recently, Z. Lin \textit{et al.}~\cite{lin2020dirac} have demonstrated massless Dirac fermions in bulk FeSn, due to the presence of a preserved combined \textit{PT} symmetry, despite the individual breaking of \textit{P} (space inversion) and \textit{T} (time-reversal) symmetries. In addition, the geometrical configuration of Fe atoms present in kagome layers shows the co-existence of both bulk and surface Dirac fermions, as well as topological flat bands~\cite{lin2020dirac, lin2018flatbands}. Recent theoretical work by J.~M.~Due\~{n}as~\cite{medina2024emerging} indicates the existence of both \textit{intrinsic} DL torque and the FL torque arising from the bulk electronic structure of Dirac semimetals. Han \textit{et al.}~\cite{han2021evidence} experimentally observed surface states and proposed that these surface states can manifest peculiar SOTs, although any experimental confirmation has yet to be reported. These promising studies suggest that interfacing a kagome antiferromagnet FeSn with a conventional ferromagnet is expected to transfer strong and unconventional SOT across the interface. This remains an unexplored territory in terms of both experiments and theory. 
Therefore, exploring the spin-orbit torques (SOTs) in a heterostructure of Dirac semimetals, FeSn, and ferromagnets such as Py is highly interesting. 

In this study, we optimize the deposition of high-quality epitaxial thin films of the antiferromagnet FeSn, utilizing a Pt seed layer on a $c$-plane Al$_2$O$_3$ substrate. 
Spin-torque ferromagnetic resonance (STFMR) measurements on FeSn/Py heterostructure reveal the presence of a conventional damping-like (DL) torque due to the $y-$spin polarization (with efficiency: $\xi_{\rm DL}^{y}$). Additionally, we observe a large unconventional field-like (FL) torque arising from the out-of-plane $z-$spin polarization (with efficiency: $\xi_{\rm FL}^{z}$). Our results reveal that the observed six-fold symmetry of $\xi_{\rm DL}^{y}$ in FeSn/Py is directly correlated with the six-fold crystal symmetry of the [0001]-oriented FeSn films. Additionally, $\xi_{\rm FL}^{z}$ displays a unique angular dependence, characterized by a combination of the six-fold symmetry inherent to the FeSn crystal structure and the uniaxial symmetry associated with the antiferromagnetic spin Hall effect. Notably, the unconventional field-like torque is enhanced when the radio frequency (RF) current is aligned with the N\'eel vector.

\section{Results and Discussion}
\subsection{Structural characterization}

Figure~\ref{Fig:1}c shows the XRD measurements performed in $\theta-2\theta$ mode for a Pt(5~nm)/FeSn(30~nm) film stack. We see strong Bragg reflections at $2\theta=(20.1^{\circ},~40.7^{\circ},~88.2^{\circ})$ corresponding to the FeSn(000\textit{l}) ($l=$2, 4, 8) planes. The presence of peaks corresponding to only (000\textit{l}) planes indicates the epitaxial growth of FeSn layer, which was later confirmed by $\phi-$scans. The additional peaks observed at $2\theta=39.8^{\circ}$, and $85.4^{\circ}$ correspond to Pt(111), and Pt(222) reflections, respectively. This confirms the epitaxial growth of the Pt layer on the $c-$plane Al$_2$O$_3$ substrate. 
In addition, we observed Laue oscillations at $2\theta=$ 34.5$^\circ$, 36.7$^\circ$, and 43.4$^\circ$, indicating a high interfacial quality of the seed layer as well as the FeSn layer~\cite{cheng2022atomic}. 

To determine the epitaxial relation between the FeSn layer, the seed layer, and the substrate, we performed $\phi-$scans on Pt(5 nm)/FeSn(30 nm) thin film stacks as shown in Fig~\ref{Fig:1}d. For the $\phi-$scan of the FeSn layer, we use the asymmetric reflection $(10\Bar{1}1)$, which is observed at a tilt angle, $\chi=45.04^\circ$ and $2\theta=27.80^\circ$. 
For the Pt layer, we use the reflection $(200)$, which is observed at $\chi=54.7^{\circ}$ and $2\theta=46.32^{\circ}$. At last, for Al$_2$O$_3$, we use the reflection ($\bar{2}116$), which is observed at  $\chi=41.86^{\circ}$ and $2\theta=57.53^{\circ}$. 
The presence of six peaks in the $\phi$-scan for Al$_2$O$_3$($\bar{2}116$) (bottom panel), Pt(200) (middle panel), and FeSn($10\bar{1}1$) (top panel) layers indicates a six-fold symmetry for both the Pt~\cite{kurosawa2024large, verguts2018growth, wardenga2016surface} and FeSn layers~\cite{khadka2020high,laxmeesha2024epitaxial}. For the Pt layer, the six-fold symmetry indicates the formation of twin domains (T$_1$, T$_2$) with 180$^{\circ}$ rotation around the Pt[111] pole with each other~\cite{kurosawa2024large}. Therefore, the in-plane epitaxial relationship corresponding to Pt T$_1$-and T$_2$-domains are Al$_2$O$_3$$[2\bar{1}\bar{1}0]\parallel$Pt$[1\bar{2}1]\parallel$FeSn$[2\bar{1}\bar{1}0$] and Al$_2$O$_3$$[2\bar{1}\bar{1}0]\parallel$Pt$[\bar{1}2\bar{1}]\parallel$FeSn$[2\bar{1}\bar{1}0$], respectively~\cite{kurosawa2024large}.
The out-of-plane crystallographic relation is found to be Al$_2$O$_3$$[0001]\parallel$Pt$[111]\parallel$FeSn$[0001]$, which agrees with previous reports~\cite{khadka2020high}. Furthermore, we performed $\omega$-scans (rocking curves) on the FeSn (0004) reflection at $2\theta = 40.81^\circ$~[Fig~\ref{Fig:1}e]. The value of the full width at half-maximum (FWHM) of the $\omega$-scan is about 0.152$^\circ$~($\approx$~547~arc-sec) for the FeSn (0004) reflection, indicating a high crystalline quality of the epitaxial FeSn layer grown on top of the Pt seed layer. The value of FWHM for our FeSn film is consistent with previous reports for a 30 nm-FeSn layer grown on top of Fe and Co seed layers~\cite{laxmeesha2024epitaxial}. 
Further, to confirm the antiferromagnetic nature of our FeSn film, we measured the exchange bias ($H_{\rm ex.}$) in a FeSn(30~nm)/Py(8~nm) bilayer, where the Py layer was deposited on top of the FeSn layer. The hysteresis loop of the ferromagnetic Py layer was initially recorded at room temperature ($T=300$~K) and subsequently at a low temperature ($T=2$~K) after cooling in the presence of a positive in-plane magnetic field of 60~kOe. At $T=2$~K, a negative shift of approximately 52~Oe in the in-plane hysteresis loop was observed for the FeSn(30~nm)/Py(8~nm) bilayer, confirming the antiferromagnetic behavior of the FeSn film. 
$H_{\rm ex}$ almost doubles ($\sim$100 Oe) on reducing the Py thickness to 5 nm confirming the interfacial nature of the exchange bias.

\begin{figure} [t!]
\centering
\includegraphics[width=\linewidth]{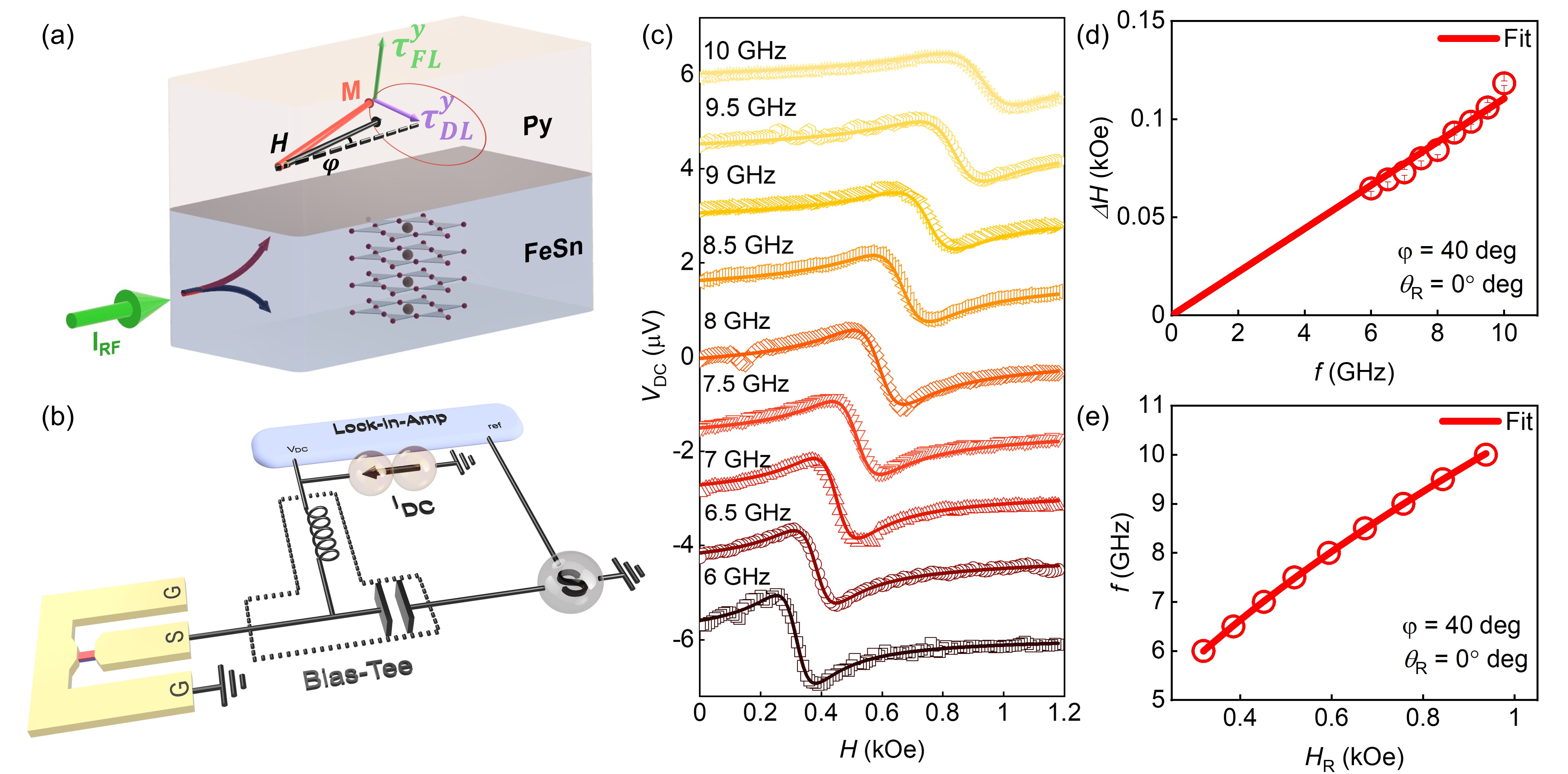}
\caption{\textbf{STFMR measurements} (a) Structure of STFMR device consisting of a FeSn/Py bilayer. This device is subject to electric current flowing along the $x-$direction, resulting in the generation of both in-plane ($\tau _{\parallel}$) and out-of-plane ($\tau _{\bot}$) torque. (b) The schematic depicts the setup for STFMR measurements. c) Frequency-dependent STFMR spectra of FeSn/Py microstrip measured at $\varphi$ = 40$^{\circ}$ when the current direction is at 0$^{\circ}$ with in-plane reference direction [1$\bar{1}$00] (The plots are shifted along the $y-$axis for clarity). (d) linewidth vs. frequency data and (e) frequency vs. resonance field data for FeSn/Py. Symbols are measured data, and solid lines are fits.
}
\label{Fig:3}
\end{figure}

\begin{figure} [t!]
\centering
\includegraphics[width=.8\linewidth]{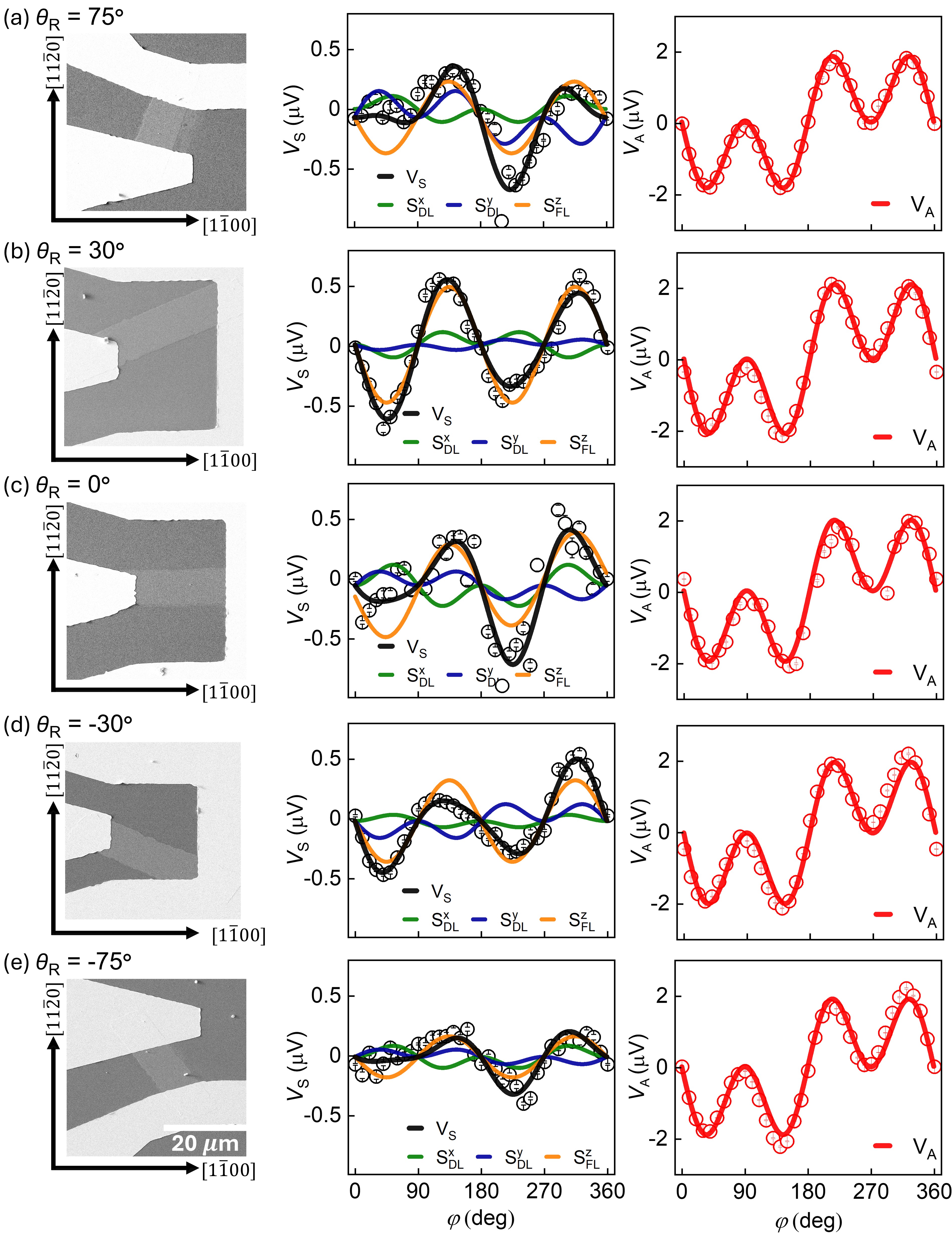}
\caption{\textbf{Crystallographic dependence of STFMR} (a-e) SEM image of differently oriented devices (left panel), $V_{\rm S}$ (middle panel) and $V_{\rm A}$ (right panel) components as a function of angle $\varphi$. $\theta_{\rm R}$ represent the angle between $I_{\rm RF}$ and in-plane reference direction [$1\bar{1}00$].}
\label{Fig:4} 
\end{figure}
\subsection{STFMR measurements on FeSn/Py}\label{sec:3.3}

To determine SOTs in the FeSn/Py bilayer, we performed STFMR measurements by passing an RF current ($I_{\rm RF}$) through the devices, which induces alternating torques on the magnetization of the ferromagnet Py, exciting its resonance, as illustrated in Fig.~\ref{Fig:3}a. We used a bias tee to apply an RF current with power +10 dBm from a signal generator, while simultaneously measuring the DC voltage. The STFMR signal was obtained by sweeping an in-plane magnetic field ($H$) through the resonance condition.
More details of the measurement setup can be found elsewhere~\cite{kumar2021large, kumar2023interfacial}.
Examples of STFMR spectra measured at different RF frequencies for $\varphi=~40$$^{\circ}$ (the angle between $H$ and $I_{\rm RF}$) are shown in Fig.~\ref{Fig:3}b. The angle, $\varphi$, was varied from 0$^{\circ}$ to 360$^{\circ}$ using a vector field magnet. Fits to the STFMR signal ($V_{\rm DC}$) were performed using a combination of symmetric and antisymmetric Lorentzian functions:~\cite{Liu2011}

\begin{equation}\label{Vmix}
 {V_{\rm DC}} = {V_{\rm S}} \frac{\Delta H^2}{\Delta H^2+(H-H_{\rm R})^2}\\+{V_{\rm A}} \frac{\Delta H(H-H_{\rm R})}{\Delta H^2+(H-H_{\rm R})^2}.
\end{equation}

Here, $V_{\rm S}$ and $V_{\rm A}$ correspond to the magnitudes of the symmetric and anti-symmetric components of $V_{\rm DC}$, proportional to the in-plane ($\tau _{\parallel}$) and out-of-plane ($\tau _{\bot}$) torques, respectively. In systems with conventional SHE and REE, 
the torques $\tau_{\parallel}$ and $\tau_{\bot}$ take the forms of $\hat m \times (\hat m \times \hat{y})$ and $\hat m \times \hat{y}$, respectively. Here, $\hat m$ and $\hat{y}$ represent the direction of the Py magnetization and the polarization of the spin current, respectively. Figure~\ref{Fig:3}c shows the linewidth ($\Delta H$) vs. RF frequency ($f$), fitted using $\Delta H=\frac{2\pi\alpha}{\gamma}f+\Delta H_0$  to obtain the Gilbert damping constant, $\alpha = 0.039 \pm 0.002$, and the inhomogeneous broadening, $\Delta H_0$~\cite{mudgal2023magnetic, gupta2024generation}. 
The strong enhancement in damping constant is due to a significant spin pumping from Py into kagome antiferromagnet FeSn~\cite{kacho2025fesn}. The inhomogenous broadening, $\Delta H_0$, was negligible, indicating a high film quality. Figure~\ref{Fig:3}d illustrates the relationship between the frequency and the resonance field, from which the effective magnetization ($\mu_{\rm 0}M_{\rm eff}$) was extracted to (1.06 $\pm$ 0.01)~T by fitting the data to the Kittel formula ~\cite{Liu2011, Macneil2017NP_WTe2}.  This value indicates that there is no significant difference in magnetic properties of the Py layers grown on FeSn.

We then analyze the results in more detail by performing complete angular dependence of the STFMR. 
In systems with conventional SHE and REE,
the angular dependence of both the symmetric ($V_{\rm S}$) and anti-symmetric voltage components ($V_{\rm A}$) of the rectified voltage follow a $\rm sin(2\varphi)sin(\varphi)$ behavior. Since FeSn is epitaxial, the STFMR measurements are carried out in FeSn/Py devices for $I_{\rm RF}$ along different angles ($\theta_{\rm R}$) with respect to the Al$_{2}$O$_{3}$ crystal structure in-plane reference [1$\bar{1}$00] direction: $\theta_{\rm R}$ = 75$^{\circ}$, 30$^{\circ}$, 0$^{\circ}$, -30$^{\circ}$ and -75$^{\circ}$ as shown in the left panel optical images of Fig.~\ref{Fig:4}. The middle and right panels of Fig.~\ref{Fig:4} show the angular dependence for the symmetric (black) and antisymmetric (red) STFMR signals, respectively. The antisymmetric component clearly follows the conventional $\rm sin(2\varphi) cos(\varphi)$-dependence. In contrast, the symmetric component differs significantly from the conventional $\rm sin(2\varphi) cos(\varphi)$-behavior~\cite{Liu2011, Macneil2017NP_WTe2}. Furthermore, the behavior changes dramatically with $\theta_{\rm R}$. To account for this deviation, extra terms proportional to $\rm sin(2\varphi)sin(\varphi)$ and $\rm sin(2\varphi)$ are introduced, corresponding to spin polarizations in the $x-$ and $z-$directions, respectively. The angular dependence of $V_{\rm S}$ and $V_{\rm A}$ for the reference sample showing the expected conventional $\rm sin(2\varphi) cos(\varphi)$-dependence. The angular dependence of $V_{\rm S}$ and $V_{\rm A}$ are fitted using~\cite{Macneil2017NP_WTe2,nan2020controlling,bose2022tilted}:

\begin{equation}\label{VS}
{V_{\rm S}} \propto {\rm sin(2\varphi)} \Bigl(S_{\rm DL}^{x}{\rm sin(\varphi)}+ S_{\rm DL}^{y}{\rm cos(\varphi)}+ S_{\rm FL}^{z}\Bigl)
\end{equation}
\begin{equation}\label{VA}
{V_{\rm A}} \propto A_{\rm FL}^{y}{\rm sin(2\varphi)}{\rm cos(\varphi)}
\end{equation}

 Here, $S_{\rm DL}^{\rm x}\propto \hat{m} \times (\hat{m} \times \hat{x})$ and $S_{\rm DL}^{\rm y}\propto \hat{m} \times (\hat{m} \times \hat{y})$, both 
 proportional to the DL torque component of the torque conductivity tensor,
 represent the unconventional component due to the $\hat{x}$ spin polarization and the conventional component due to the $\hat{y}$ spin polarization, respectively. Similarly, $A_{\rm FL}^{\rm y}\propto (\hat{m} \times \hat{y})$ and $S_{\rm FL}^{\rm z}\propto (\hat{m} \times \hat{z})$,
 represent the conventional component due to the
 $\hat{y}$ spin polarization and the unconventional component due to 
 the $\hat{z}$ spin polarization, respectively. The superscript refers to the direction of the spin polarization for a charge current applied along $\hat{x}$. 

By fitting the angular dependence of $V_{S}$ and $V_{A}$ with Eq.~\ref{VS} and Eq.~\ref{VA}, the different torque components can be separated, with their efficiencies 
expressed as~\cite{zhou2020magnetic,bose2022tilted}:
\begin{equation}\label{eqSHA_x}
\xi_{\rm DL}^{x} = \frac{S_{DL}^{x}}{A_{FL}^{y}}\frac{e\mu_{0}M_{\rm S}t_{\rm FM}d_{\rm NM}}{\hbar}\left[1+\frac{4\pi M_{\rm eff}}{H_{\rm R}}\right]^{1/2}
\end{equation}

\begin{equation}\label{eqSHA_y}
\xi_{\rm DL}^{y} = \frac{S_{DL}^{y}}{A_{FL}^{y}}\frac{e\mu_{0}M_{\rm S}t_{\rm FM}d_{\rm NM}}{\hbar}\left[1+\frac{4\pi M_{\rm eff}}{H_{\rm R}}\right]^{1/2}
\end{equation}

\begin{equation}\label{eqSHA_z}
\xi_{\rm FL}^{z} = \frac{S_{FL}^{z}}{A_{FL}^{y}}\frac{e\mu_{0}M_{\rm S}t_{\rm FM}d_{\rm NM}}{\hbar}\left[1+\frac{4\pi M_{\rm eff}}{H_{\rm R}}\right]^{1/2}
\end{equation}

Here, $d_{\rm NM}$ and $t_{\rm FM}$ are the non-magnetic and ferromagnetic layer thicknesses, respectively, and 
$e$ the electric charge. The values of $M_{\rm eff}$, $H{\rm _R}$ are calculated from the frequency-dependent STFMR measurements, as discussed above.

\begin{figure} [t!]
\centering
\includegraphics[width=0.8\linewidth]{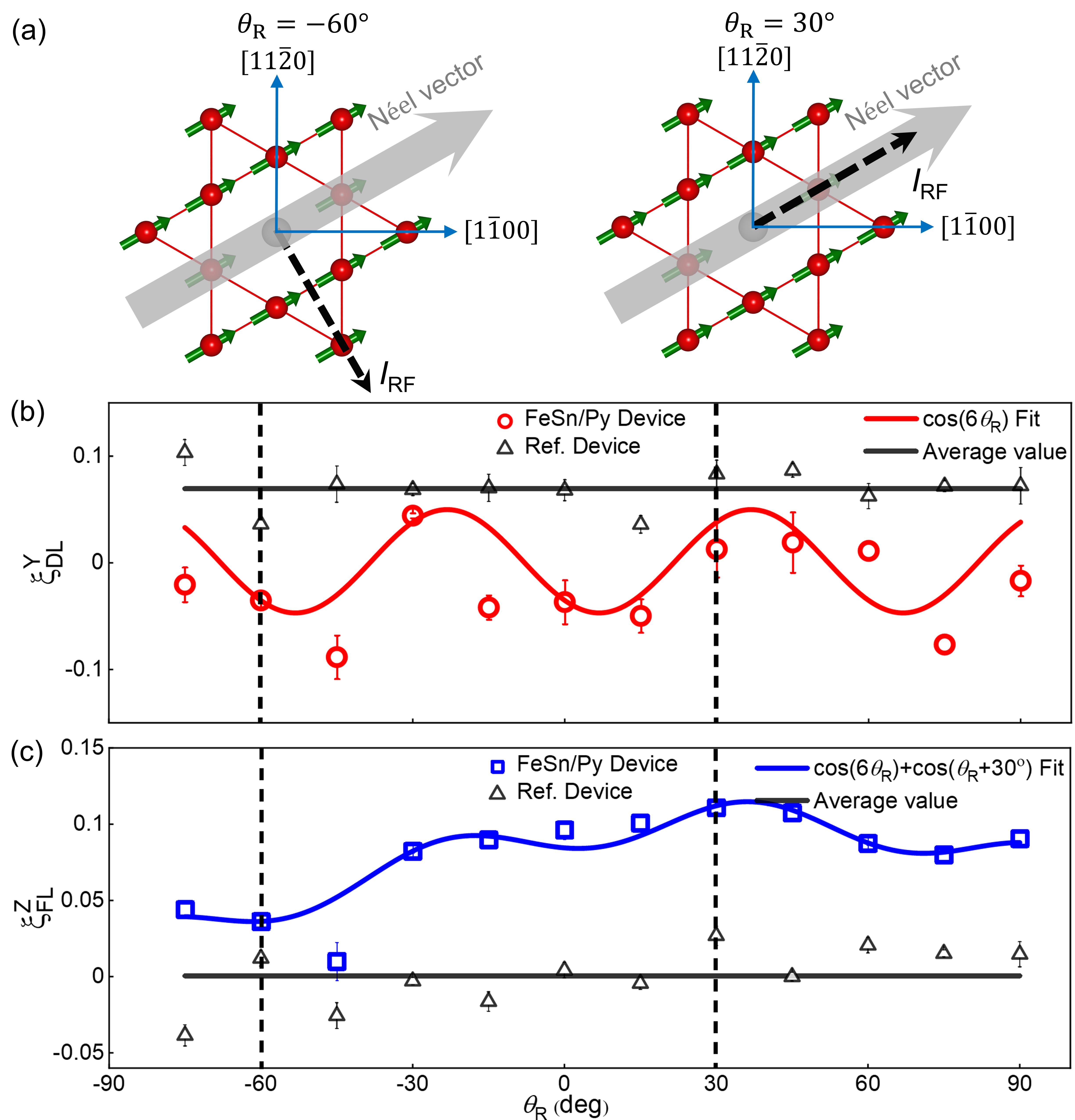}
\caption{\textbf{Crystallographic dependence of spin-torque efficiencies} (a) Kagome Structure with N\'eel vector (gray) and I$_{RF}$ (black dashed) in parallel and perpendicular orientations (b) DL SOT efficiency due to $y-$ spin polarization and (c) FL SOT efficiency due to $z-$ spin polarization for different $\theta_{\rm R}$ values in FeSn/Py device and Ref. device.}
\label{Fig:5}
\end{figure}
\section{Crystallographic dependence of spin-torque efficiencies}

Figure~\ref{Fig:5}a shows the kagome structure with $\theta_{\rm R}$ = -60$^{\circ}$ and 30$^{\circ}$, where the RF current is perpendicular and parallel to the N\'eel vector, respectively. The top and bottom panels in Fig.~\ref{Fig:5}b\&c show the measured DL and FL SOT efficiencies, respectively, as a function of $\theta_{\rm R}$ in FeSn/Py and Ref. devices. The conventional DL SOT efficiencies due to the $y-$ spin polarization, denoted as $\xi_{\rm DL}^{y}$, are shown in  Fig.~\ref{Fig:5}b for the Py/FeSn device and for the Ref. device. For the Ref. device, $\xi_{\rm DL}^{y}$ is independent of $\theta_{\rm R}$,
which is expected for the isotropic SHE of Pt
~\cite{gudin2023isotropic, guillemard2018charge}. In contrast, for the FeSn/Py heterostructure in Fig.~\ref{Fig:5}b, $\xi_{\rm DL}^{y}$ exhibits a strong anisotropic behavior, and 
even sign inversions for different $\theta_{\rm R}$ (different crystal axes), 
which are entirely absent in the Ref. device. The $\theta_{\rm R}$ dependence of $\xi_{\rm DL}^{y}$ can be fitted using a $\cos(6\theta_{\rm R})$ function, illustrated by the red solid line in Fig.~\ref{Fig:5}b. This fit indicates a six-fold symmetry in the SOT response, consistent with the underlying six-fold symmetry of the epitaxial [0001]-oriented FeSn films as observed in the $\phi$-scan measurements [Fig.~\ref{Fig:1}d]. Thus, $\xi_{\rm DL}^{y}$ can be linked to the bulk electronic properties of the FeSn kagome structure.
 
We also observe a large unconventional FL SOT due to the spin current with out-of-plane, or $z-$polarization, labeled as $\xi_\mathrm{FL}^\mathrm{z}$, as shown in Fig.\ref{Fig:5}c. As expected, this unconventional torque, $\xi_\mathrm{FL}^\mathrm{z}$, is essentially absent in the Ref. device. In contrast, for the FeSn/Py heterostructure, $\xi_\mathrm{FL}^\mathrm{z}$ is not only large but also exhibits a strong anisotropic behavior, though there is no sign inversion of $\xi_\mathrm{FL}^\mathrm{z}$ with $\theta_{\rm R}$  unlike the case of $\xi_{\rm DL}^{y}$.
The $\theta_{\rm R}$ dependence of $\xi_{\rm FL}^{z}$ can be modeled using a $\cos(6\theta_{\rm R})+\cos(\theta_{\rm R}+30$$^{\circ}$) function, which is illustrated by the blue solid line in Fig.~\ref{Fig:5}c. This fit indicates a combination of a six-fold symmetry and a uniaxial symmetry in the SOT response. The six-fold symmetry is consistent with the six-fold symmetry of the epitaxial [0001]-oriented FeSn films as observed in $\xi_{\rm DL}^{y}$. However, the uniaxial term with a shift of 30$^{\circ}$~indicates the presence of an additional mechanism for the generation of out-of-plane spin current.

\section{Discussion}

In the following, we discuss the potential origin of the torques observed in FeSn/Py-based devices. FeSn is an antiferromagnetic kagome metal that exhibits exotic surface states. Han \textit{et al.}~\cite{han2021evidence} experimentally observed surface states and proposed that these surface states can manifest peculiar SOTs.  
Additionally, FeSn is a Dirac semimetal, and recent theoretical work by J.~M.~Due\~{n}as~\cite{medina2024emerging} indicates the existence of both \textit{intrinsic} DL torque and the FL torque arising from the bulk electronic structure of Dirac semimetals. The intrinsic torques originating from the band structure inherently possess anisotropic characteristics. Consequently, these torques are expected to exhibit a strong dependence on the crystallographic direction. The conventional DL torque, $\xi_\mathrm{DL}^\mathrm{y}$  has strong anisotropic behavior with six-fold symmetry, consistent with X-ray measurements of the symmetry of epitaxial [0001]-oriented FeSn films. Therefore, the conventional DL torque, $\xi_{\rm DL}^{y}$, can be attributed to an intrinsic origin arising from the topological features of the electronic band structure of FeSn. The presence of a six-fold component in the unconventional FL torque, $\xi_{\rm FL}^{z}$, is also consistent with the above theoretical work~\cite{medina2024emerging}, which predicts simultaneous enhancement of DL and FL torques.

The $\theta_{\rm R}$ dependence of $\xi_{\rm FL}^{z}$ has an additional $\cos(\theta_{\rm R}+30^{\circ}$) component. Our results suggest that $\xi_{\rm FL}^{z}$ also requires the presence of a FeSn layer adjacent to Py. Several recent studies have demonstrated the occurrence of an out-of-plane polarization resulting from the magnetic spin Hall effect in non-collinear antiferromagnets~\cite{zhang2017strong,vzelezny2017spin,kimata2019magnetic,mook2020origin}.
Although FeSn is classified as a collinear antiferromagnet—which typically would not generate such torques—a recent investigation revealed the antiferromagnetic spin Hall effect in the collinear antiferromagnet, Mn$_2$Au~\cite{chen2021observation}. They reported a FL torque arising from out-of-plane spin polarization when the current direction is parallel to the N\'eel vector. For FeSn, the N\'eel vector lies along the $a-$axis~\cite{sankar2024experimental}, as shown in Fig.~\ref{Fig:5}a, which corresponds to $\theta_{\rm R}$ = 30$^{\circ}$. 
Therefore, we anticipate that for the antiferromagnetic spin Hall effect in FeSn, $\xi_{\rm FL}^{z}$  will be maximized at 
$\theta_{\rm R}$ = 30$^{\circ}$. Indeed, as shown in Fig.~\ref{Fig:5}c, there is an observed increase in 
$\xi_{\rm FL}^{z}$  by over 50\% $\theta_{\rm R}$ = 30$^{\circ}$~compared to 
$\theta_{\rm R}$ = -60$^{\circ}$, where the RF current is perpendicular to the N\'eel vector. This observation supports the conclusion that the uniaxial $\cos(\theta_{\rm R})$ component can be attributed to the antiferromagnetic spin Hall effect. While the exact mechanism of the DL torque due to $x-$ polarization $\xi_{\rm DL}^{x}$, is not known, we believe it is also related to the interface between FeSn and Py since it does not exhibit a clear six-fold symmetry or uniaxial symmetry with $\theta_{\rm R}$.

\section{Conclusion}

In conclusion, we demonstrate successful epitaxial growth of high-quality FeSn thin films on $c$-plane Al$_2$O$_3$ substrate utilizing a Pt seed layer. 
A comprehensive investigation of spin-orbit torques (SOTs) in FeSn/Py bilayers as a function of crystallographic orientation reveals a six-fold symmetric DL SOT and the coexistence of six-fold symmetric and uniaxial unconventional FL torques. The latter is attributed to the antiferromagnetic spin Hall effect, highlighting the unique potential of FeSn for SOT applications. 
The results highlight the unique properties of FeSn as a platform for exploring emergent SOT phenomena. By leveraging the topological features of FeSn and its interaction with adjacent ferromagnetic layers, all grown using industrially compatible co-sputtering, we anticipate that our results will inspire further exploration into antiferromagnetic spintronics relevant for direct SOT applications.

\section{Experimental Details }

\subsection{Sample preparation}
We prepared high-quality Pt(5~nm)/FeSn(30~nm)/Py(8~nm) multilayer film on $c$-plane Al$_2$O$_3$(0001) substrates using magnetron sputtering. The Pt layer's high quality was achieved with a growth rate of 0.27~${\mathring{\mathrm{A}}~\rm s^{-1}}$ (RF power of 60 W) and a substrate temperature of 400~$^\circ {\rm C}$. The FeSn layer was grown using the co-sputtering technique. The sputtering rates of Fe and Sn targets were optimized by adjusting the power of each sputter gun to achieve the desired stoichiometry. Using 40~W of DC power for Fe and 85~W of RF power for Sn, we obtained a low net deposition rate of 0.4 ~${ \mathring{\mathrm{A}}~\rm s^{-1}}$ for the FeSn film. The FeSn layer was grown at a substrate temperature of 550~$^\circ {\rm C}$. The ferromagnetic Py layer (8~nm) was deposited at room temperature with a low growth rate of 0.25~${\mathring{\mathrm{A}}~\rm s^{-1}}$ using 40 W of DC power. Additionally, we grew a reference sample (Ref): Pt(5~nm)/Py(10~nm) on $c$-plane Al$_2$O$_3$(0001) substrates (without FeSn layer) under the same growth conditions. 
Both samples were capped with an Al layer deposited at a growth rate of 0.20~${\mathring{\mathrm{A}}~\rm s^{-1}}$ (RF power of 60 W). The base pressure of the sputtering chamber was better than 6.7$\times 10^{-8}$~mbar, and the working pressure was maintained at 6.7$\times 10^{-3}$~mbar. The substrate holder was rotated at 60~rpm during deposition to ensure better homogeneity of the films.
After deposition, these samples have been patterned into microstrip devices of dimensions 40 $\mu$m long and 10 $\mu$m wide prepared by the lift-off technique using optical lithography.

\subsection{Structural characterization}
The actual stoichiometry of our Fe and Sn in FeSn thin films was found to be 50.2$\pm$0.5~at.\% and 49.7$\pm$0.6~at.\% by averaging the value at four different spots on the sample using the electronic probe microscopy analysis (EPMA) method.
The $\theta-2\theta$ (Gonio-XRD) and $\phi-$scan XRD measurements have been used to investigate the epitaxial nature of the films. The high-resolution XRD measurements were performed using a PANalytical X$'$Pert diffractometer with Cu-$K_\alpha$ radiation ($\lambda= 1.5418 ~\mathring{\mathrm{A}}$). 

\subsection{STFMR mesurements}

We conducted STFMR measurements by applying an RF current using R$\&$S signal generator (SMB 100A) with frequency, $f$ at a constant RF power of $+$10~dBm to the device. The microstrip is connected via a Ground-signal-Ground probe. The RF signal was amplitude-modulated with a 98~Hz signal, and the resulting DC mixing voltage was detected with a lock-in amplifier (SR830). A bias tee is utilized to separate the DC and RF port connected to the microstrip. An angle-dependent spin-torque ferromagnetic resonance (STFMR) was performed using vector field magnet to quantify the spin-orbit torque in the FeSn/Py system. The magnetic field was applied in the sample plane at an angle $\varphi$ with respect to the current direction ($x-$direction) along the long axis of the microstrip. The angle, $\varphi$, was varied from $(0-360^{\circ})$ with the help of a vector field magnet (GMW 5201).

\section{Declarations}

The partial support from the Ministry of Human Resource Development under the IMPRINT program (Grants No. 7519 and No. 7058), the Department of Electronics
and Information Technology (DeitY), the Science and Engineering Research Board (SERB File No. CRG/2018/001012 and No. CRG/2022/002821), Joint Advanced Technology Centre at IIT Delhi, Grand Challenge Project, IIT Delhi, and the Department of Science and Technology under the Nanomission program [grant no: $SR/NM/NT-1041/2016(G)$], IEEE Magnetics Society Special Project [P.O. 502182] are gratefully acknowledged. We also acknowledge the Central Research Facility, IIT Delhi, for providing facilities for sample characterization.

\section*{Conflict of interest}
The authors declare that there is no conflict of interest with respect to this paper. 

\bibliography{Main}

\end{document}